\documentclass[twoside]{article}
\usepackage{fleqn,espcrc2,amssymb}

\usepackage{graphicx}

\newcommand{\AmS}{{\protect\the\textfont2
  A\kern-.1667em\lower.5ex\hbox{M}\kern-.125emS}}
\title{3D-melting features of the irreversibility line in overdoped
Bi$_2$Sr$_2$CuO$_6$ at ultra-low temperature and high magnetic
field}

\author{A. Morello\address{High Magnetic Field Laboratory, Max Planck Institut f\"ur
Festk\"orperforschung and Centre National de la Recherche
Scientifique, BP166, 38042 Grenoble Cedex 9, France}\address{INFM
- Dipartimento di Fisica, Politecnico di Torino, c.so Duca degli
Abruzzi 24, 10129 Torino, Italy}%
        \thanks{Present address: Kamerlingh Onnes Laboratory - Leiden University,
         P.O. Box 9504, NL 2300 RA Leiden, The Netherlands.},
         A.G.M. Jansen$^{\rm a}$,
         R.S. Gonnelli$^{\rm b}$
         and
         S.I. Vedeneev\address{ P.N. Lebedev Physical Institute, Russian Academy of Sciences,
         SU-117924 Moscow, Russia}}

\begin{document}

\begin{abstract}
We have measured the irreversible magnetization of an overdoped
Bi$_2$Sr$_2$CuO$_6$ single crystal up to $B=28$ T and down to
$T=60$ mK, and extracted the irreversibility line $B_{\rm
irr}(T)$: the data can be interpreted in the whole temperature
range as a 3D-anisotropic vortex lattice melting line with
Lindemann number $c_{\rm L}=0.13$. We also briefly discuss the
applicability of alternative models such as 2D- and quantum
melting, and the connection with magnetoresistance experiments.
\vspace{1pc}
\end{abstract}

\maketitle

\section{INTRODUCTION}

Despite the deep structural and physical similarities with the
wide family of Bi- and Tl- based high-$T_{\rm c}$ compounds, the
layered cuprate superconductor Bi$_2$Sr$_2$CuO$_6$ (Bi-2201) has a
relatively low critical temperature (typically less than 13 K). On
the other hand, this is accompanied by a value of the upper
critical field $B_{\rm c2}$ which lies within the experimentally
accessible range, thus allowing the investigation of the whole
$B-T$ phase diagram. Nevertheless, difficulties in growing
high-quality single crystals have made this material much less
studied than the other cuprates. In this work we present the
results of irreversible magnetization measurements, carried out up
to the applied magnetic field $B_{\rm a}=28$ T and down to $T=60$
mK on a Bi-2201 single crystal. The goal of this work is to
extract the irreversibility line $B_{\rm irr}(T)$ and deduce what
is the most suitable model to interpret its behavior, especially
in view of the high anisotropy of the material and the influence
of flux motion in the resistive transitions in magnetic field.

\section{EXPERIMENT}

The investigated sample is a high-quality
Bi$_{2+x}$Sr$_{2-(x+y)}$Cu$_{1+y}$O$_{6\pm \delta }$ single
crystal of approximate size $1100\times 700\times 10$ ${\rm
\mu}$m$^3$, intrinsically overdoped because of the Bi excess
localized on the Sr positions; accordingly, we found a critical
temperature $T_{\rm c}\approx 4$ K. The magnetization was detected
by means of a very sensitive capacitive torquemeter, installed
into the mixing chamber of a dilution refrigerator. The plane of
the torquemeter was tilted in order to have an angle
$\theta=30^\circ$ between the $c$-axis of the sample and the
direction of the applied magnetic field. The torque loops $\tau
(B)$ (inset of Fig. 1) were recorded during two independent sets
of experiments, sweeping the field at the rates ${\rm d}B_{\rm
a}/{\rm d}t=10.8$ mT/s and ${\rm d}B_{\rm a}/{\rm d}t=15$ mT/s,
respectively. The magnetization loops $M(B)$ can be obtained from
the relationship $M=\tau /(B_{{\rm a}}\sin \theta)$. Calling
$B_{\rm irr}^{\rm (a)}$ the applied field corresponding to the
vanishing of the irreversible torque, the actual irreversibility
field $B_{\rm irr}$ that would be obtained for $\bf{B}_{\rm
a}\parallel c$ is given by $B_{\rm irr}=B_{\rm irr}^{\rm (a)} \cos
\theta$; this is due to the high anisotropy of Bi-2201, implying
that only the field orthogonal to the $ab$-planes is effective.

\begin{figure}
\includegraphics[keepaspectratio,width=7.5cm]{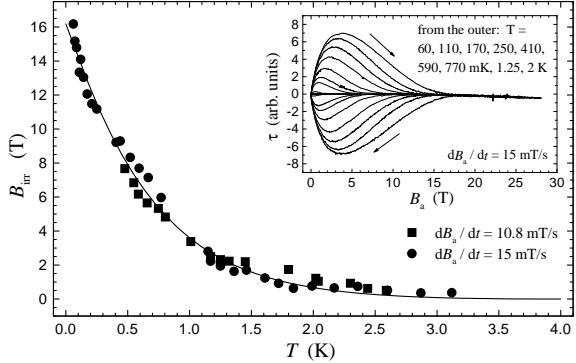}
\caption{The irreversibility line of our Bi-2201 sample can be
fitted to the 3D-anisotropic vortex lattice melting line as
calculated from the Lindemann criterion, yielding $c_{\rm L}=0.13$
(solid line). The inset shows that all the torque loops have a
typical shape that only scales with temperature.} 
\end{figure}

\section{DISCUSSION}

The irreversibility line $B_{\rm irr}(T)$ obtained in this way is
shown in Fig. 1: qualitatively, it is very similar to the
resistive upper critical field recently obtained with crystals
very similar to ours \cite{vede99}, provided that the criterion
$\rho /\rho_{\rm n}=0.1$ (i.e. the "foot" of the resistive
transition) is chosen. This confirms that, in very anisotropic
superconductors, there is a wide area of the $B-T$ phase diagram
where the resistive transition is influenced by flux motion. On
the other hand, this $B_{\rm irr}(T)$ can not be interpreted in a
simple depinning or flux creep picture, since its curvature does
not correspond to a law $(1-T/T_{\rm c})^n$ with $n \leqslant{2}$.
Here we show that the flux motion above $B_{\rm irr}$ takes place
as a consequence of the melting of a 3D-anisotropic flux lattice,
as demonstrated by the very good fit (solid line in Fig. 1)
provided by the analytical form of the 3D-melting line $B_{{\rm
m}}(T)$ as calculated from the Lindemann criterion \cite{blat94}:

\begin{equation}
B_{{\rm m}}(T)=B_{{\rm c2}}(0)\frac{4\vartheta ^2}{\left( 1+\sqrt{%
1+4\vartheta T_{{\rm s}}/T}\right) ^2}
\end{equation}

where $\vartheta =c_{{\rm L}}^2\sqrt{\beta _{{\rm m}}/Gi}(T_{{\rm
c}}/T-1)$,
$T_{{\rm s}}=T_{{\rm c}}c_{{\rm L}}^2\sqrt{\beta _{{\rm m}}/Gi}$, $c_{{\rm L}%
}$ is the Lindemann number, $Gi=\frac 12\left( \frac{\gamma k_{{\rm B}}T_{%
{\rm c}}}{(4\pi /\mu _0)B_{{\rm c}}^2(0)\xi_{ab} ^3(0)}\right) ^2$
is the Ginzburg number, $\gamma =\sqrt{m_c/m_{ab}}$ is the mass
anisotropy, $\kappa =\lambda _{ab}(0)/\xi _{ab}(0)$ and $\beta
_{{\rm m}}\approx 5.6$. With parameters appropriate for Bi-2201,
the fit to Eq. (1) yields the very reasonable estimate of $c_{\rm
L}=0.13$, very close to the value $c_{\rm L}=0.14$ predicted for
3D vortex lattice melting in high magnetic field \cite{hika91}. It
is interesting to observe that Eq. (1) is obtained using a linear
extrapolation of $B_{{\rm c2}}$ down to $T=0$, and here it
provides a good fit down to $T/T_{\rm c}\sim 0.01$: a fully linear
behavior of $B_{{\rm c2}}(T)$ is indeed found in Bi-2201
\cite{vede99} taking the saturation points of the magnetoresistive
transitions.\\ The high anisotropy of Bi-2201 might suggest that a
more realistic picture for the vortex system is a stack of weakly
coupled 2D pancake vortices \cite{glaz91}. Instead, our data do
not agree neither qualitatively nor quantitatively to the
predicted existence of a 2D-3D crossover field $B_{{\rm cr} }$ and
a field-independent melting temperature ${T} _{{\rm m} }^{{\rm
2D}}$ in the $B-T$ phase diagram; also the almost identical shape
of the torque loops at different temperatures (inset of Fig. 1)
suggests the existence of only one pinning mechanism. We can also
rule out a possible quantum melting of the vortex lattice at low
temperatures \cite{blat93}, since we don't find any low-$T$ linear
features in $B_{\rm irr}(T)$.\\ In conclusion, we can interpret
the irreversibility line of overdoped
Bi$_{2+x}$Sr$_{2-(x+y)}$Cu$_{1+y}$O$_{6\pm \delta }$ as the
signature of 3D-anisotropic vortex lattice melting down to
$T/T_{\rm c}\sim 0.01$.

\end{document}